\documentclass[prl,twocolumn,notitlepage,showpacs, superscriptaddress,aps]{revtex4-1}
\usepackage{graphicx}
\usepackage{amsmath}
\usepackage{amssymb}
\newcommand{\ham}{{\mathcal{H}}}
\newcommand{\carb}{$^{13}$C }
\newcommand{\ku}{\vert{0}\rangle}
\newcommand{\bu}{\langle{0}\vert}
\newcommand{\ket}[1]{\left\vert{#1}\right\rangle}
\newcommand{\bra}[1]{\left\langle{#1}\right\vert}
\newcommand{\sx}{\sigma_x}
\newcommand{\sy}{\sigma_y}

\newcommand{\tr}[1]{\textrm{Tr}\left\{{#1}\right\}}
\usepackage{multirow}
\usepackage{hhline}
\usepackage[usenames,dvipsnames]{color} 
\definecolor{LinkColor}{rgb}{0,0,.5}
\usepackage[colorlinks=true,citecolor=LinkColor,urlcolor=LinkColor]{hyperref}

\renewcommand{\emph}{\textit}

\graphicspath{{/Users/pcappell/Documents/Work/Papers/OptimalDecomp/OC-paper-June/}}

\begin{document}
\title{The NV center as a quantum actuator: time-optimal control of nuclear spins}
\author{Clarice D.~Aiello}
\address{Department of Electrical Engineering and Computer Science,}
\author{Paola Cappellaro}\email{pcappell@mit.edu}
\address{Department of Nuclear Science and Engineering and Research
Laboratory of Electronics,
Massachusetts Institute of Technology, Cambridge, MA, USA}
\begin{abstract}
Indirect control of qubits by a quantum actuator has been proposed as an appealing strategy to manipulate qubits that couple only weakly to external fields.  While universal quantum control can be easily achieved when the actuator-qubit coupling is anisotropic, the efficiency of this approach is less clear. Here we analyze the time-efficiency of the quantum actuator control. We describe a strategy to find time-optimal control sequence by the quantum actuator and  compare their gate times  with direct driving, identifying regimes where the actuator control performs faster. As an example, we focus on a specific implementation based on the Nitrogen-Vacancy center electronic spin in diamond (the actuator) and nearby carbon-13 nuclear spins (the qubits). 
\end{abstract}
\maketitle
\paragraph*{\textbf{Introduction --}}\label{sec:Intro}
Fast and high fidelity control of quantum systems is a key ingredient for quantum computation and sensing devices. The critical task is to reliably control a quantum system, while staving off decoherence,   
by keeping it isolated from any external influence. These requirements pose a contradiction: fast control implies a strong coupling to an external controlling system, but this entails an undesired interaction with the environment, leading to decoherence. 
One is then often faced with the choice  between using a strongly connected  system, implying a stronger noise, or a weakly connected one, which is more isolated from the environment and thus offers longer coherence times, but results in slower control.  
A strategy to overcome these issues is to use a hybrid system where a quantum actuator interfaces the quantum system of interest to the classical controller, thus allowing fast operations while preserving the system isolation and coherence. This indirect control is particularly appropriate for nuclear spin qubits, which only couple weakly to external fields, but often show strong interactions with nearby electronic spins. This model describes several systems, from spins associated with phosphorus donors in Si~\cite{Morton08}, to fullurene qubits~\cite{Morton06}, ensemble-ESR systems (such as malonic acid~\cite{Hodges08}) and, most recently, Nitrogen-Vacancy (NV) centers in diamond~\cite{Cappellaro09,Taminiau14}. While there are practical advantages to this indirect control strategy, as it does not require experimental apparatus to directly drive the nuclear spins, an important question is whether it can reach faster manipulation than direct control. In this letter we describe a strategy to achieve time-optimal indirect control of a nuclear spin qubit by an electronic spin quantum actuator; for the specific case of the NV center in diamond, we assess the achievable gate times of this strategy as compared with those for direct driving. 

We propose to use alternating controls to drive the evolution of a nuclear spin anisotropically hyperfine-coupled to an electronic spin~\cite{Hodges08,Khaneja07}; in particular, periodically driving the spin of a NV center in diamond can steer the evolution of a proximal $^{13}$C nuclear spin in a potentially shorter time than a direct, slow radio-frequency (rf) addressing. In general, the method ensures the use of the nuclear spin as a resource within the same implementation time range of direct addressing, while entirely by-passing the application of rf, thus avoiding any noise associated with it~\cite{Gerstein}. 

\paragraph*{\textbf{Alternating Control --}}\label{sec:Alternate}
Universal control of a target qubit by a quantum actuator can be achieved as long as the two systems present an anisotropic coupling. Consider a single NV center coupled to a \carb nuclear spin. Their Hamiltonian is
\[\ham=\Delta S_z^2+\gamma_eB_0S_z+\gamma_CB_0I_z+\vec S\cdot A\cdot \vec{I} \ ,\]
where $\Delta=2.87$GHz is the NV zero-field splitting; $\gamma_{e}\!\approx\! 2.8$MHz/G, $\gamma_C\!\approx\!1$kHz/G are, respectively, the gyromagnetic ratios of the electron and nuclear spins; $B_0$ is a static magnetic field along the NV $\hat{z}$-axis; and $A$ is the hyperfine tensor. This can be rewritten in the electronic spin rotating frame, when driving the $|m_s= 0\rangle\leftrightarrow|\pm1\rangle$ transition, and neglecting the off-resonant manifold, as
\begin{equation}
\ham\!=\!\omega_0 I_z+S_z \vec{A}_z \cdot \vec I\!=\!\ku\!\bu \omega_0I_z+\ket{\pm1}\!\bra{\pm1}(\omega_0I_z\pm\vec{A}_z \cdot \vec I) \ ,
\end{equation}
where $\omega_0= \gamma_CB_0$ (that we assume $>0$). The  contact and dipolar contributions to the hyperfine coupling $\vec{A}$ can be described by a longitudinal component $A$ and a transverse component $B$, that we will take  without loss of generality along the $\hat{x}$ direction~\cite{Schweiger01}. The nuclear spin thus evolves by rotating around two distinct axes, depending on the electronic spin manifold. Concatenating rotations about these two axes is enough to achieve full controllability of the nuclear spin~\cite{Hodges08}. Then, a simple strategy for the indirect control of the nuclear spin is to induce alternating rotations by flipping the electronic spin state with (fast) $\pi$-pulses.  

We define the axes and rotation speeds in the two manifolds as
\begin{equation}\begin{array}{ll}
\omega_0=\gamma_C B_0=\kappa\omega_{\pm1} \ ,\ \ \,& \omega_{\pm1}=\sqrt{(\omega_0\pm A)^2+B^2} \ ,\\  \hat v_0=\hat z\ ,& \hat v_1=\hat z\cos(\alpha)+\hat x\sin(\alpha) \ ,
\end{array}
\end{equation}
with \vspace{-12pt}
\begin{equation}
\tan(\alpha)=\frac{B}{\omega_0\pm A}\ ,\qquad \kappa=\frac{\omega_0}{\omega_{\pm1}} \ .
\end{equation}

If the NV  electronic spin is initially in the $\ket{0}$ state, by applying $\pi$-pulses at times $T_k$, the nuclear spin evolution is given by
\begin{equation}
U=\dots e^{-i\phi_k\vec{v}_0\cdot\vec\sigma}e^{-i\phi_{k-1}\vec{v}_1\cdot\vec\sigma}\dots e^{-i\phi_i\vec v_0\cdot\vec\sigma},
\end{equation}
where $\phi_k=(T_k-T_{k-1})\omega_{0(1)}$, for odd (even) $k$, and $\vec\sigma$ are the Pauli matrices.

In order to compare the actuator-control strategy with direct driving, we need to consider the time-optimal way to synthesize  the desired unitary $U$ by alternating rotations~\cite{Billig13,Boscain04b,Boscain06}. This problem has been recently addressed using algebraic methods~\cite{Billig13,Aiello14} and we only describe here the most important results, relevant to the problem at hand. 
\paragraph*{\textbf{Time-optimal control --}}
In order to find time-optimal sequences of rotations, a greatly simplifying condition is that the solution depends on only four parameters: the outer angles $\phi_i, \phi_f$, the internal angle $\phi_0$ of the rotation around $\vec v_0$, and the total number of rotations $n \leq \infty$ (the sequence length). If $n\geq4$, the internal angles are related by
\begin{equation}
\tan\left(\frac{\phi_1}{2}\right) = \tan\left(\frac{\phi_0}{2}\right)  \frac{\kappa - \cos(\alpha)}{1 - \kappa \cos(\alpha)} \ ,  
\label{eq:angles}
\end{equation}
with $\phi_1$ the rotation angle about $\vec v_1$.
We can distinguish two important cases that yield different time-optimal solutions, whether $\kappa \lessgtr \cos(\alpha)$. This condition is simply set by the sign of the longitudinal hyperfine interaction (and the chosen manifold), since it corresponds to $\omega_0 \lessgtr(\omega_0\pm A) $.

If $\kappa<\cos(\alpha)$, optimal sequences are finite and we always have $\phi_0\leq\pi$ and $\phi_1\geq\pi$. Finite sequences with $n\geq6$ have $\pi/3<\phi_0<\pi$ and their length is bound by $n\leq \lfloor{\frac{2\pi}{\alpha}}\rfloor+1$.

For $\kappa > \cos(\alpha)$, both finite and infinite time-optimal sequences are possible. For large angles between rotation axes, $\alpha>2\pi/3$, only $n=3$ or infinite sequences are possible, with $\phi_0>\pi$. For smaller angles, we can have longer time-optimal sequences. 
The number of switches is limited by $n\leq \lfloor\frac{\pi}{\alpha}\rfloor + 3$ 
and, correspondingly, we have $\pi < \phi_0 \leq \frac{(n-1)}{(n-2)}\pi$. Loose bounds can also be found for the outer angles~\cite{Aiello14} and thus on the total time to implement general unitaries.

These conditions on the admissible time-optimal sequences severely constraint the search space of the time-optimal control sequence for specific goal unitaries and Hamiltonian parameters. We were thus able to perform an exhaustive analysis of time-optimal control for a large number of nuclear spins surrounding the NV center.
\paragraph*{\textbf{Direct driving of nuclear spins --}} While the nuclear spin coupling to an external driving is weak, indirect forbidden transitions mediated by the electronic spin can considerably enhance the driving strength~\cite{AbragamBleaney}. This nuclear Rabi enhancement  depends on the state of the electronic spin. The effective Rabi frequency $\Omega$ for an isolated nuclear spin, hence, is modified from its bare value $\overline\Omega$ by the enhancement factors $\zeta_{0,\pm 1}$ (corresponding to the electronic spin states $|0\rangle, |\pm 1\rangle$). To first order in the electronic spin detuning from the rf frequency we have
\begin{equation}\begin{array}{l}
\zeta_0-1 \!\!=\!  - \frac{\gamma_e}{\gamma_n}\frac{4  {B} (\Delta-A)}{ (\Delta + B_0 (\gamma_e - \gamma_n)-A) (\Delta - B_0 (\gamma_e - \gamma_n)-A)} ; \\
=- \frac{4 B_0 \gamma_e\sin(\alpha) (\kappa  (B_0 \gamma_n+\Delta )- B_0 \gamma_n \cos(\alpha))}
{[\kappa  (B_0 \gamma_e+\Delta )-B_0 \gamma_n \cos(\alpha)] \{\kappa[\Delta-B_0 (\gamma_e -2  \gamma_n) ]-B_0 \gamma_n \cos(\alpha)\}}\\
\zeta_{+1}\!-\!1 \!\!=\! \frac{\gamma_e}{\gamma_n}\frac{2 {B}}{\Delta + B_0 (\gamma_e - \gamma_n)-A }=\frac{2 B_0 \gamma_e\sin(\alpha)}{\kappa  (B_0 \gamma_e+\Delta )-B_0 \gamma_n \cos(\alpha)}  ; \\
\zeta_{-1}\!-\!1 \!\!=\!  \frac{\gamma_e}{\gamma_n} \frac{2B}{ \Delta - B_0 (\gamma_e - \gamma_n)-A}= \frac{2 B_0 \gamma_e\sin(\alpha)}{\kappa[\Delta-B_0 (\gamma_e -2  \gamma_n) ]-B_0 \gamma_n \cos(\alpha) }
\ .
\end{array}
\label{eq:enhancement}
\end{equation}
Note that $\zeta_i$ can be either positive or negative, and $\lessgtr 1$. Since $\gamma_e/\gamma_n\!\approx\!2600$, the effective Rabi frequencies $\Omega_i=\zeta_i\overline\Omega$ can be much larger than the bare frequency.

We assume $\overline\Omega \!\approx\! 100$kHz as an upper-limit on realistic bare nuclear Rabi frequencies by considering data in~\cite{Maurer12}, where the $^{13}$C considered was only weakly coupled and thus no Rabi enhancement was present. To achieve this strong driving, a dedicated microfabricated coil was necessary~\cite{Sun13}. We note that in some instances, for high enhancement factors, a limitation to the permissible Rabi frequency is further set by the desire of not breaking the Rotating Wave Approximation~\cite{Bloch40}.
Rabi frequencies $\overline\Omega \!\approx\! 20$kHz are, in our experience, in the upper achievable range with modest amplifiers and a simple wire to deliver the rf field.

\paragraph*{\textbf{Comparison of control schemes --}}
Both regimes $\kappa \lessgtr \cos(\alpha)$ can be explored considering the coupling to $^{13}$C at different distances from the NV defect~\cite{Gali08, Felton09,Dreau12}. The hyperfine tensors for $^{13}$C located up to $\!\approx\! 8$\AA \ away from the NV center were estimated using density functional theory~\cite{Smeltzer11,Galiprivate}. In what follows, we numerically compare the performance of the proposed control method against direct driving under diverse experimental conditions and for a number of distinct nuclear spins. 

Using the relationship between internal angles given by Eq.~\ref{eq:angles} and the bounds on their values,  we numerically searched for sequences $U^\star$. 
The search was deemed successful when the fidelity $F \equiv \frac{1}{2}|\textrm{tr}(U^\star  U_{\scriptsize{\textrm{goal}}}^\dagger)| = 1-\epsilon$, with $\epsilon\lesssim 10^{-10}$. We repeat the search and choose the sequence with minimal time cost among all sequences obtained in successful searches to ensure that we are at the global time-optimum within numerical error. 

\begin{figure}[t]
\centering
 \includegraphics[width=0.21\textwidth]{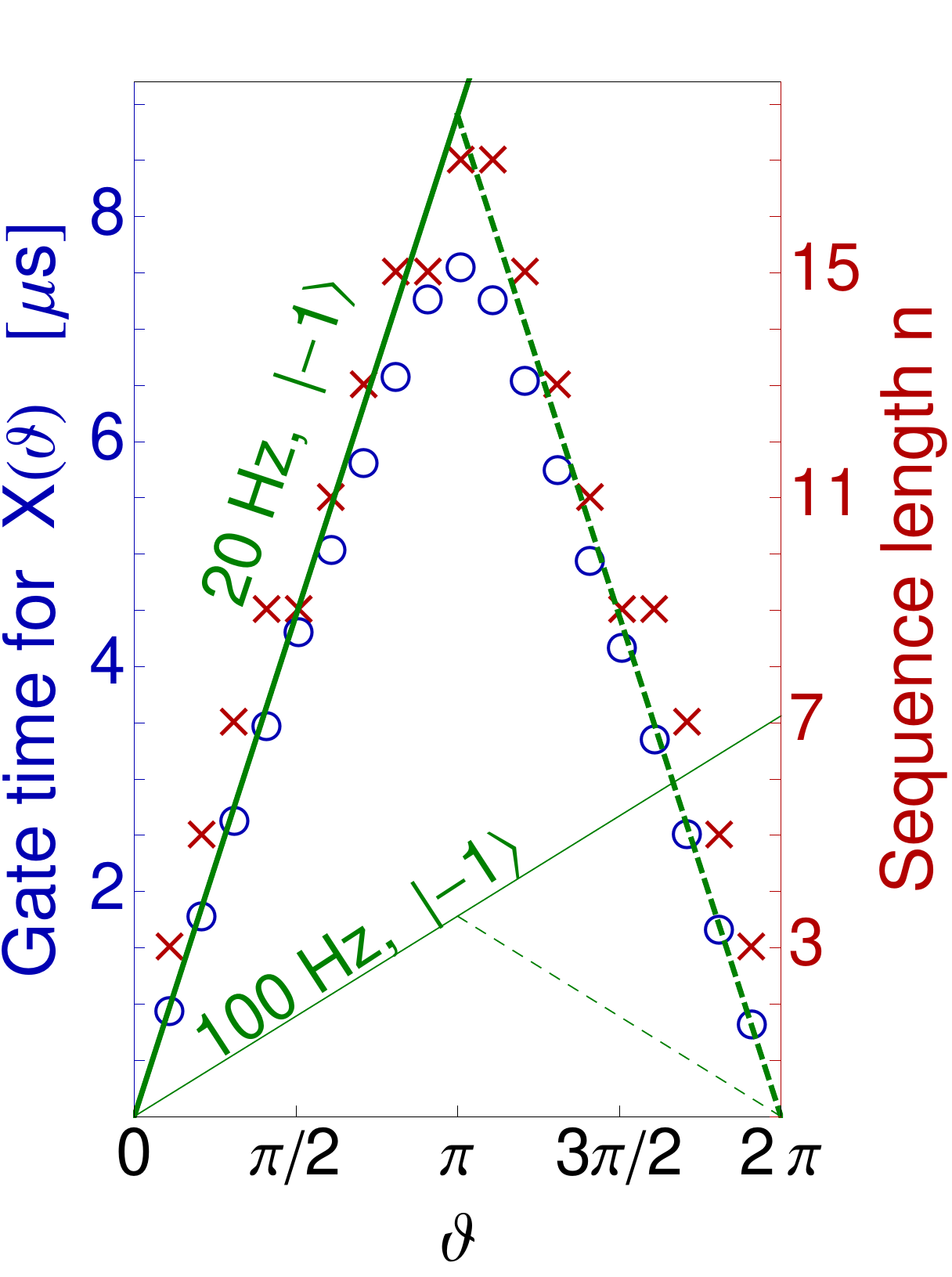}\qquad 
  \includegraphics[width=0.21\textwidth]{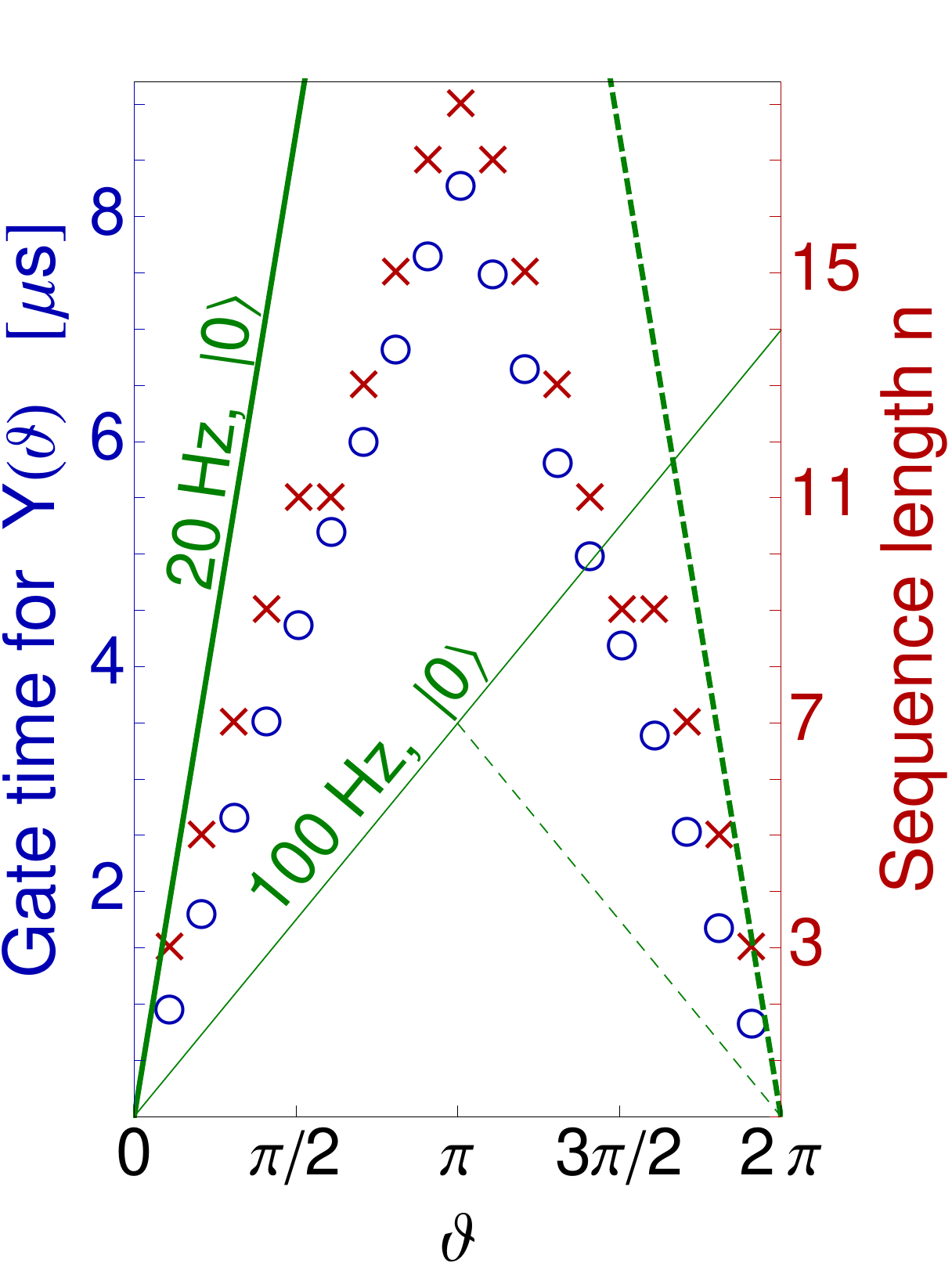} 
\caption{\textbf{Comparison of gate time: Case $\kappa < \cos(\alpha)$,}  occurring for a $^{13}$C at a distance of $\!\approx\! 2.92$\AA \ from the NV center, with an external magnetic field $B_0 \!\approx\! 500$G  aligned with the $\hat{z}$ axis. 
We plot the simulated actuator implementation time (blue circles-left axis) of the unitaries $\mathsf{X}(\theta)$ (left) and $\mathsf{Y}(\theta)$ (right) and the corresponding sequence lengths (red crosses-right axis). For comparison, we plot the time required with direct driving (green lines) with bare Rabi frequencies $20$ and $100$kHz,  when the electronic spin in state $|-1\rangle$ (left), thus maximizing the enhancement factor, or $\ku$ (right). Note that the direct-driving time for $\theta>\pi$ depends on whether the driving phase can be inverted (dashed line) or not (solid line).}
\label{fig:2impl}
\end{figure}

Typical results for the case $\kappa < \cos(\alpha)$ are illustrated in Figure~\ref{fig:2impl} by a $^{13}$C at a distance $r\!\approx\! 2.92$\AA \ from the NV center, at an external magnetic field $B_0\!\approx\! 500$G ($\omega_0\!=\!0.5$MHz) aligned with the $\hat{z}$ axis. 
This magnetic field strength is experimentally convenient: it achieves fast nuclear spin polarization since in the electronic excited state the nuclear and electronic spins have similar energies, allowing polarization transfer during optical illumination. We will consider later the effects of different magnetic field strengths.
The hyperfine interaction of this spin, ${A} \!\approx\! 1.98$MHz and ${B} \!\approx\! 0.51$MHz, yields $\alpha \!\approx\! 11.6^{o}$ and  $\kappa \!\approx\! 0.20$. {Although the upper bound on the sequence length is 32, we found that the optimal sequences were much shorter (red crosses).}
The simulation results indicate that, given a rotation angle $\theta$, the actuator implementation times for rotations around any axis in the $\{\hat{y}, \hat{x}\}$ plane are comparable, with a maximum around $\theta \!\approx\! \pi$, and a symmetry for $\theta =\pi\pm \delta$. 
We plot, in particular, the optimal times $T_A(\theta)$ required to generate the unitaries $\mathsf{X}(\theta)\equiv e^{-i\theta\sx/2}$ and $\mathsf{Y}(\theta)\equiv  e^{-i\theta\sy/2}$ with the actuator scheme (blue circles). Here and in the following we neglect the time needed for the actuator $\pi$-pulses, since it can be as low as 2-5ns~\cite{Fuchs09}. 
For comparison, we consider direct driving with bare Rabi frequencies in the range $\overline\Omega \!\approx\! 20-100$kHz. In Figure~\ref{fig:2impl} we plot the gate time  $T_D(\theta)$ required with directive driving (green solid and dashed lines), taking into account the Rabi enhancement factors, which for this nuclear spin are  $\zeta_0 \!\approx\! -1.43$, $\zeta_{+1} \!\approx\! 1.62$ and $\zeta_{-1} \!\approx\! 2.81$.
Note that for bare Rabi frequencies weaker than $\!\approx\! 20$kHz, the actuator protocol is advantageous for any rotation angle. 
\begin{figure}[tb]
\centering
\includegraphics[width=0.21\textwidth]{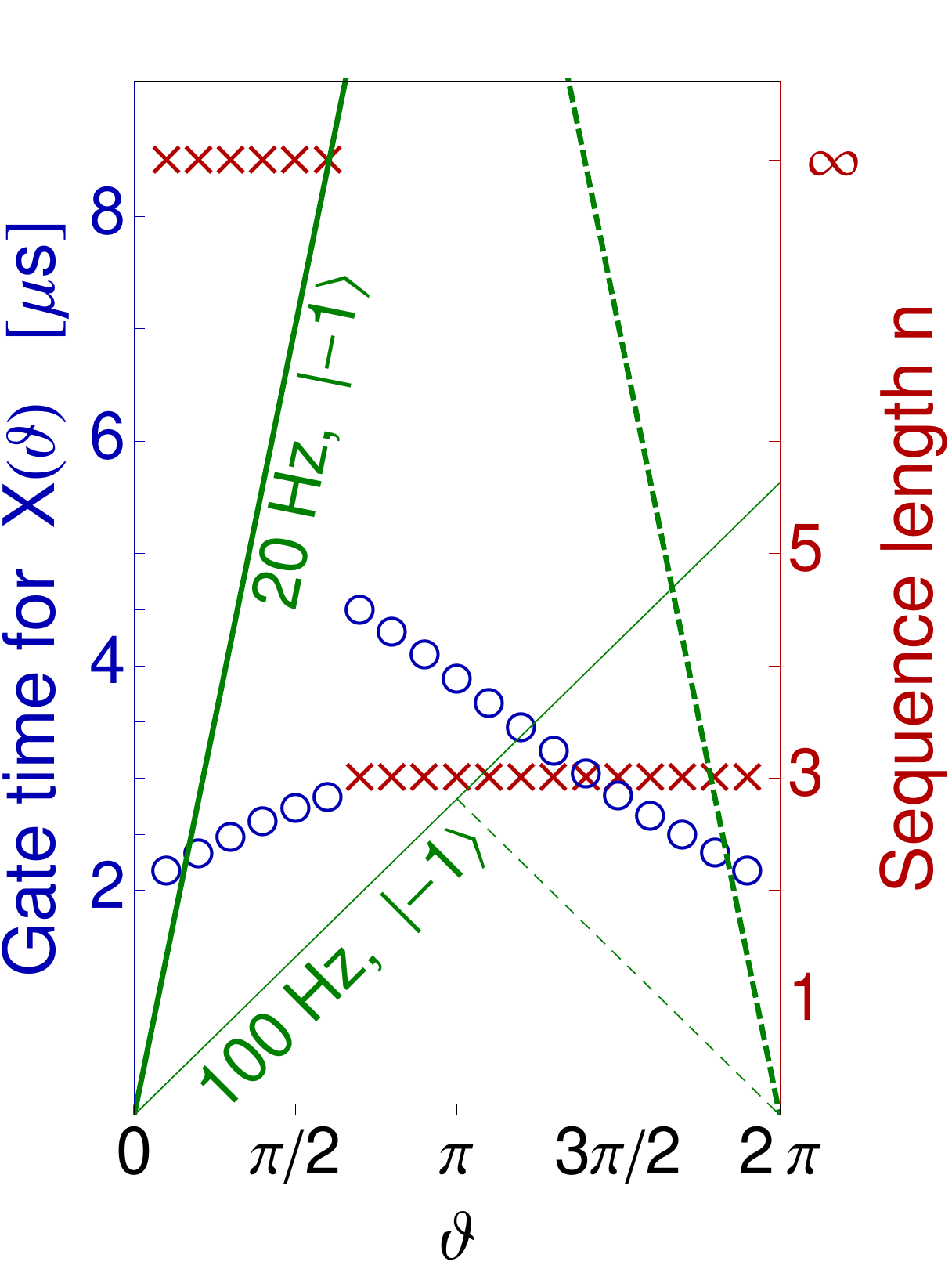} \qquad
\includegraphics[width=0.21\textwidth]{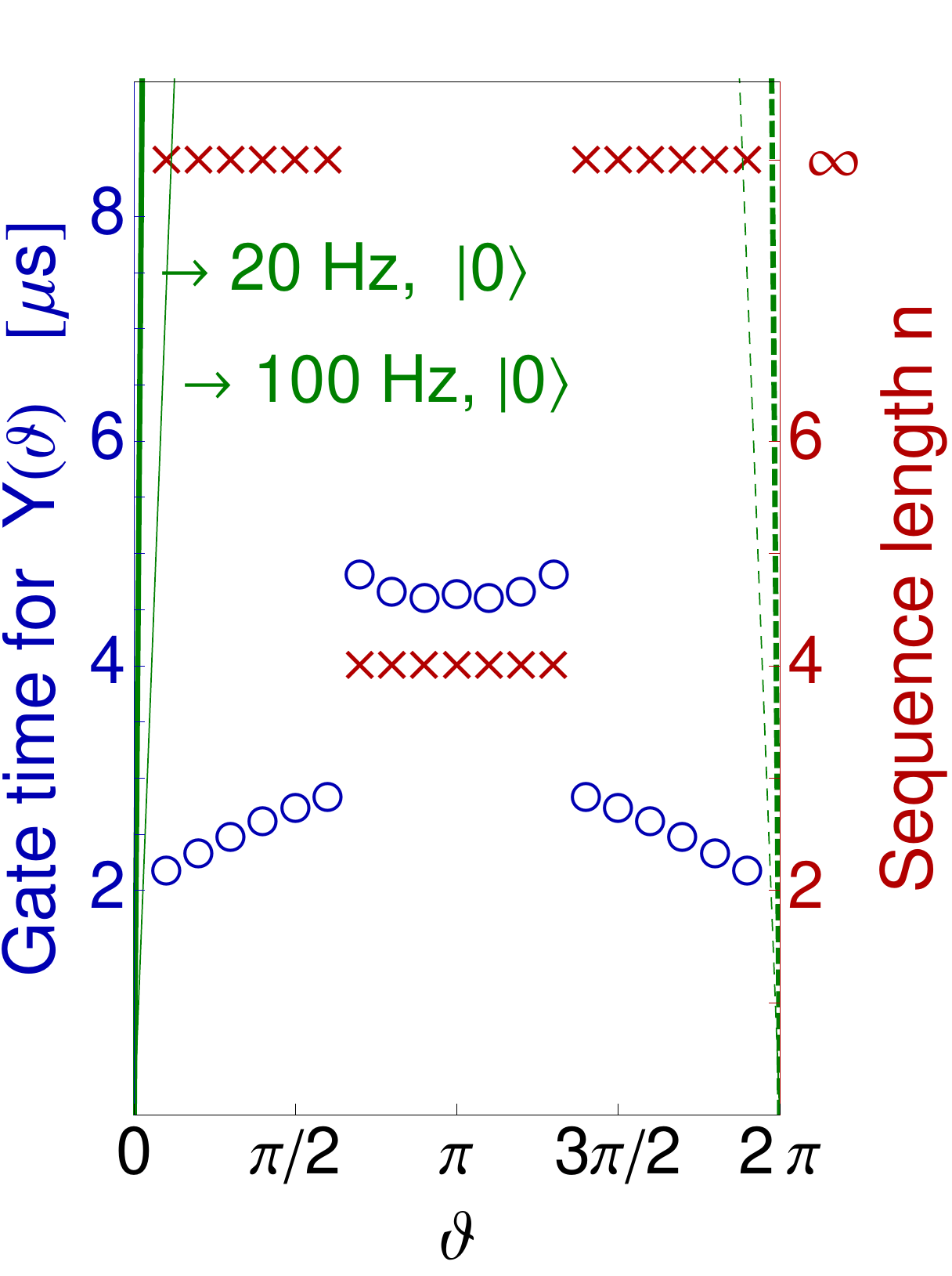} 
\caption{
\textbf{Comparison of gate time: Case $\kappa > \cos(\alpha)$,}  occurring for a $^{13}$C at a distance of $\!\approx\! 4.31$\AA \ from the NV center. See Figure~\ref{fig:2impl} for comparison and explanation of symbols. Note in particular that the Rabi enhancement in the $m_s=0$ manifold is $<1$, thus making direct driving in that manifold unfavorable.}
\label{fig:1impl}
\end{figure}

In Figure~\ref{fig:1impl}, we examine the driving of a $^{13}$C at a  distance of $\!\approx\! 4.31$\AA \ from the NV center, for which ${A} \!\approx\! -0.35$MHz and ${B} \!\approx\! 0.23$MHz. Under the same magnetic field conditions, $B_0 \!\approx\! 500$G, we have $\kappa \!\approx\! 1.8$, $\alpha \!\approx\! 57.4^{o}$, and thus $\kappa > \cos(\alpha)$, with the maximal possible length of a finite time-optimal sequence being $n = 6$. 
The figures show the optimal times to synthesize the unitaries $\mathsf{X}(\theta)$ and $\mathsf{Y}(\theta)$ as a function  of the rotation angle $\theta$ as well as the corresponding length of the  time-optimal sequence. Note that for the synthesis of some unitaries, the optimal scheme requires infinite-length sequences.
We compare the time required with the actuator protocol to the direct driving, taking into account the enhancement factors ($\zeta_0 \!\approx\! -0.07$, $\zeta_{+1} \!\approx\! 1.29$ and $\zeta_{-1} \!\approx\! 1.78$). 
Even if the hyperfine coupling strength is smaller than for the first spin considered, the actuator times are in general smaller; similarly, 
even for the highest considered direct-driving Rabi frequency the actuator protocol can have a lower time-cost.

\begin{figure}[t]\centering
\includegraphics[width=0.21\textwidth]{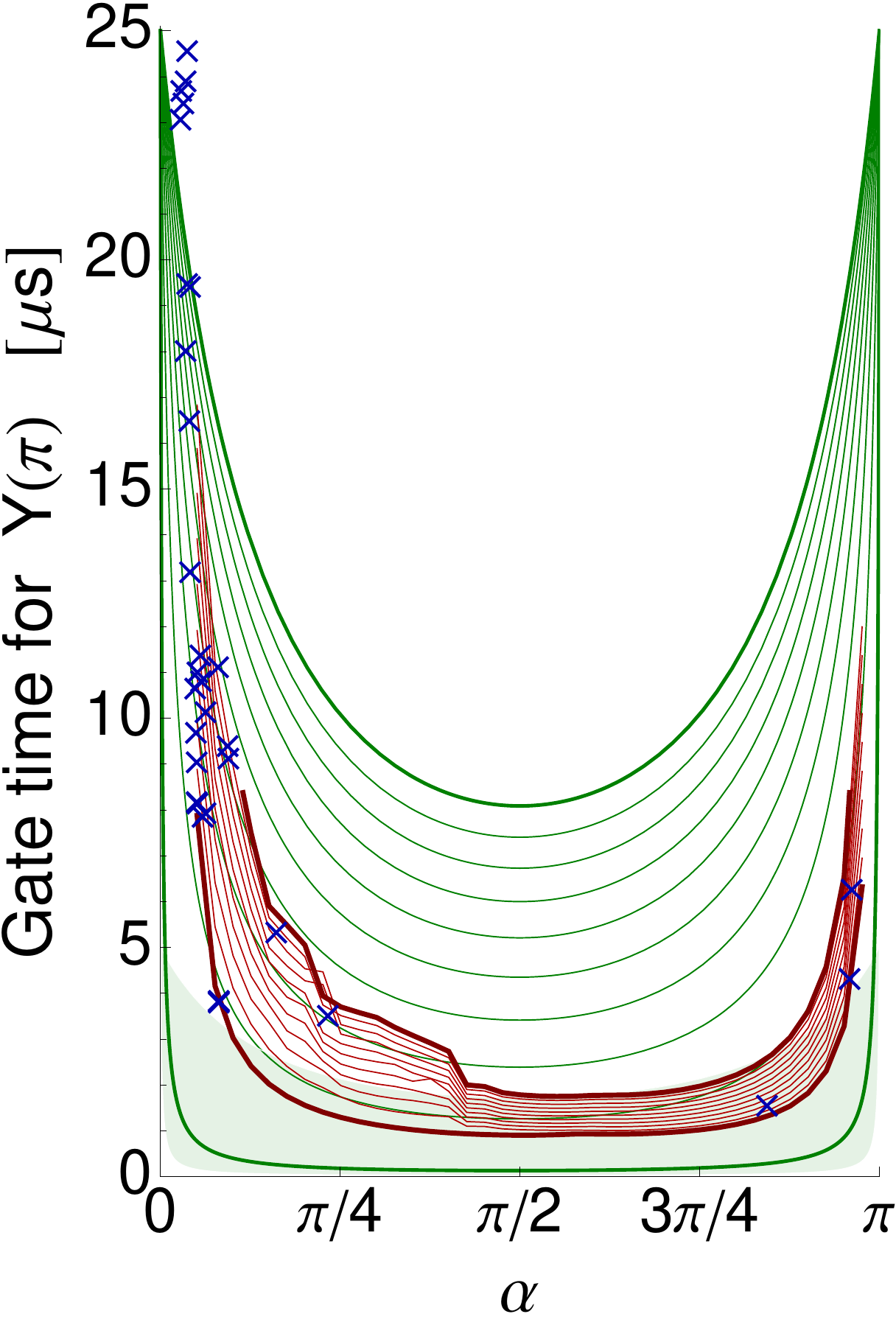} \qquad
\includegraphics[width=0.21\textwidth]{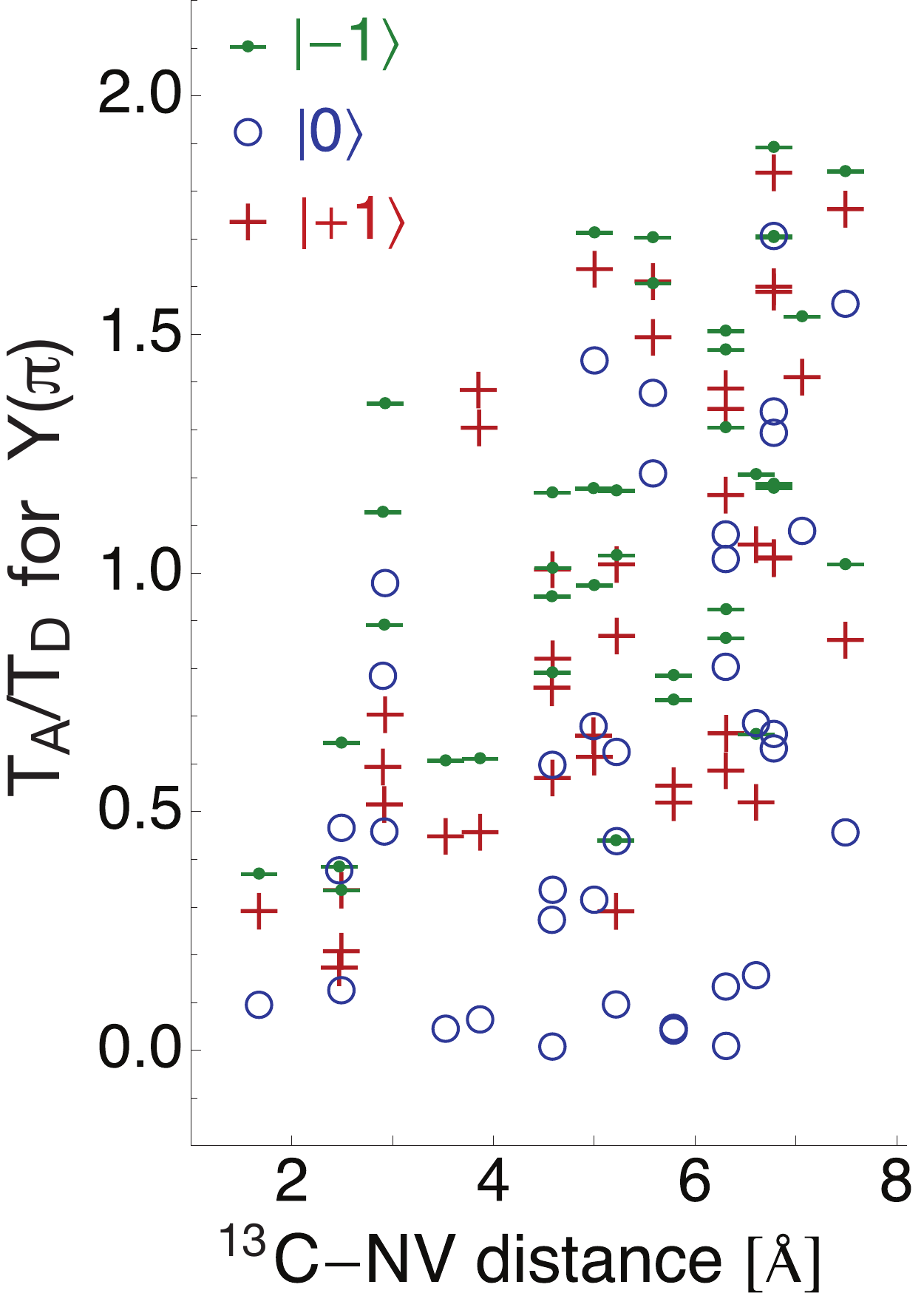} 
\caption{In the left panel, we show the simulated actuator implementation times for $\mathsf{Y}(\pi)$, for the entire $\alpha$ range. Values for $\kappa$ span from $10^{-3}$ (bottom thick red line), through $0.1$ to $0.9$ in $0.1$ intervals (thin red lines), to $1$ (top thick red line). The actuator implementation times to generate $\mathsf{Y}(\pi)$ for all the tabulated carbons are plotted for comparison using blue crosses. For the same values of $\kappa$, we plot the direct driving times for Rabi frequencies $\Omega \!\approx\! 20$kHz (green lines) and $\Omega \!\approx\! 100$kHz (green shaded region). 
In the right panel, we display the ratio of actuator to direct driving times in the generation of $\mathsf{Y}(\pi)$, for all three electronic spin states $|0\rangle, |+1\rangle, |-1\rangle$ (blue circles, red plus signs, and green dash-dot, respectively), in the case of bare Rabi frequency $\overline\Omega \!\approx\! 20$kHz, and as a function of the distance between nuclear and electronic spins.}
\label{fig:3impl}
\end{figure}

While the results shown for particular nuclear spins are indicative of the achievable gate times, the broad range of hyperfine coupling tensors can give rise to quite different behaviors.
We thus numerically investigate the actuator implementation time of a particular unitary $\mathsf{Y}(\pi)$ for an extended range in $\{\alpha, \kappa\}$ space; the result is plotted in the leftmost panel of Figure~\ref{fig:3impl}. To find the times for a smooth set of parameters,  
we interpolate  the implementation times found numerically for each tabulated nuclear spin (parametrized by its unique pair $\{\alpha, \kappa\}$). 

We compare the times achievable with the actuator scheme with the times required for direct driving. In particular, we use Eq.~\ref{eq:enhancement} to calculate effective Rabi frequencies over the same range of parameters $\{\alpha,\kappa\}$. 
If only a moderate driving strength is available (a bare Rabi frequency of $\Omega \!\approx\! 20$kHz) the actuator scheme is faster than direct driving for a broad region of the parameter space. While \carb nuclear spins coupled to the NV center do not span the whole region, other systems might, presenting an even more favorable situation. 

As shown in Figure~\ref{fig:3impl} (right panel), for the NV center system the dependence on the hyperfine parameters of both the actuator scheme time and the direct driving strength yields a broad variation of results for both close-by and more far away nuclei; while a trend toward longer times for the actuator scheme vs. direct driving is apparent as the distance from the NV center increases, the large variations indicate that the best scheme should be evaluated for individual nuclear spins.

\begin{figure}[t]
\centering
\includegraphics[width=0.35\textwidth]{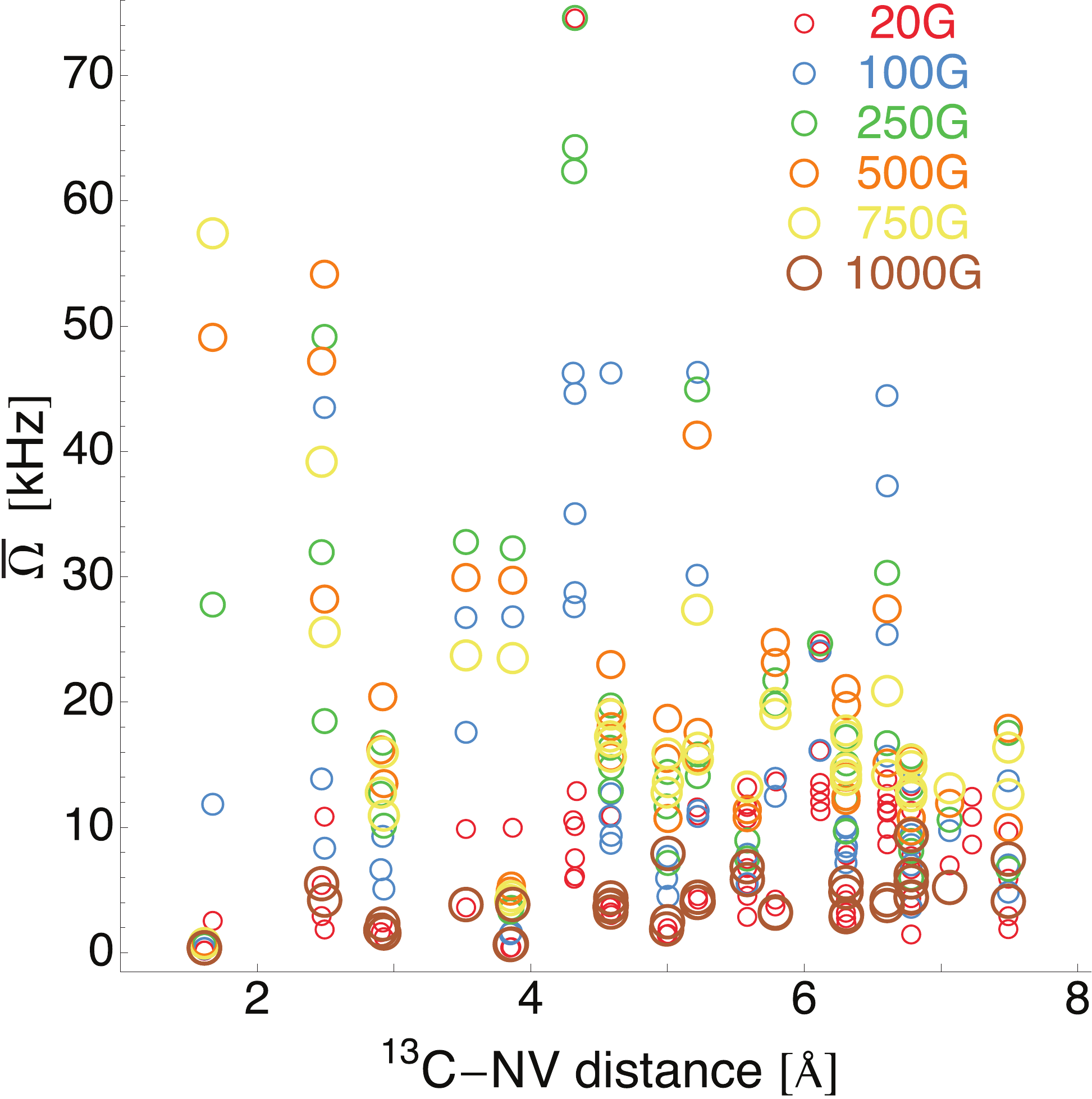} 
\caption{Minimal bare Rabi frequency for which direct driving is advantageous over the actuator method for the implementation of $\mathsf{Y}(\pi)$, for different magnetic fields.}
\label{fig:4impl}
\end{figure}

Finally, we analyze the effect of the external magnetic field strength. As it increases, the angle $\alpha$ between the two axes of rotation decreases and thus we expect longer sequences (both in terms of number of switches and of total time). On the opposite end, if the magnetic field is small, the rotation speeds decrease in both manifolds; thus, although the time-optimal sequences might have short lengths, the total time could still be long. Again, variations in the hyperfine coupling parameters yield broad variations on top of this expected behavior, as shown in Figure~\ref{fig:4impl}. For different values of the external magnetic field, we plot the bare Rabi frequency for which the actuator implementation time of $\mathsf{Y}(\pi)$ coincides with the minimum direct driving time (that is, when the enhancement factor is maximal), as a function of the  distance of the hyperfine-coupled $^{13}$C spins from the NV. 
If the available experimental bare Rabi frequency is lower than the depicted value at any given field, the actuator control method  will yield an advantage over direct driving. At intermediate fields, around $B_0 \!\approx\! 250-500$G, Rabi frequencies that favor direct driving are relatively large, indicating a region where actuator control can prove especially beneficial.

Note, incidentally, that the upper bound on the implementation time of any considered unitary, $T \!\approx\! 25\mu$s, is still much shorter than the nitrogen-vacancy center spin-lattice relaxation time at room temperature, $T_1 \!\approx\!$ 1-10ms~\cite{Jarmola12}.

\paragraph*{\textbf{Discussion -- }}
Indirect control of qubits by a quantum actuator is an attractive strategy in many situations when the qubits couple weakly to external fields but interact more strongly to another quantum system. Here we analyzed an exemplary situation, consisting of a hybrid quantum register composed of electronic and nuclear spins centered around the NV center in diamond. Using this particular system, we analyzed the parameter space where indirect control by an actuator presents a time-advantage over direct control methods. The comparison was performed by using time-optimal control results. Similar control schemes have been proposed and experimentally implemented previously, as it was realized early on that switched control is universal~\cite{Hodges08,Khaneja07}; however, time-optimality was not considered. For example, the most frequent scheme~\cite{Taminiau14,Liu13,Borneman12} applies alternate rotations for equal times; even if this is a convenient way of implementing dynamical decoupling on the actuator while manipulating the qubits, the scheme is not time-optimal and has in general poor fidelity except in the limit of small qubit/actuator coupling~\cite{SOM}. In contrast, here the electronic spin was used just as an actuator (always in a population state), and as such dynamical decoupling is not required.

An interesting extension of our results would be to \textit{simultaneously} control two or more qubits by the same quantum actuator. While this is possible, provided the qubits are coupled with different strengths~\cite{Hodges08,Schirmer01}, it becomes more difficult to find time-optimal solutions except for particular tasks (such as state-to-state transformations~\cite{Assemat10}) or geometries~\cite{Burgarth10,Zhang11}.
Still, even when the goal is to control a larger number of qubits, our results can guide the experimentalist's choice between direct driving and the actuator control, for which these results give an upper bound.

\begin{acknowledgments}
It is a pleasure to thank Boerge Hemmerling,  Michele Allegra, Xiaoting Wang and Seth Lloyd for stimulating discussions. We thank Adam Gali for providing us the hyperfine coupling strengths.This work was supported in part by the U.S. Air Force Office of Scientific Research through the Young Investigator Program grant No. FA9550-12-1-0292. C.D.A acknowledges support from Schlumberger.
\end{acknowledgments}

\bibliographystyle{apsrev4-1}
\bibliography{/Users/pcappell/Documents/Work/Papers/Biblio}

\newpage
\begin{widetext}
\appendix
\section{Supplementary Figures}
As discussed in the main text, there is a great variety in the properties of nuclear spin qubits in diamond. Here we survey some of the relevant properties for the comparison of direct driving versus the actuator model. We considered the \carb nuclear spin in the first 5 lattice cells around the NV center. As shown in Figure~\ref{fig:RabiEnh}, there is a great variation in the hyperfine parameters, even for spins that are located at similar distances from the NV center. This in turns translates into a spread in the enhancement factors of the Rabi driving frequency (left panel) and the magnitude and angle of the axis of rotation in the $m_s=1$ manifold (right panel).
\begin{figure}[h]
\centering
\includegraphics[width=0.35\textwidth]{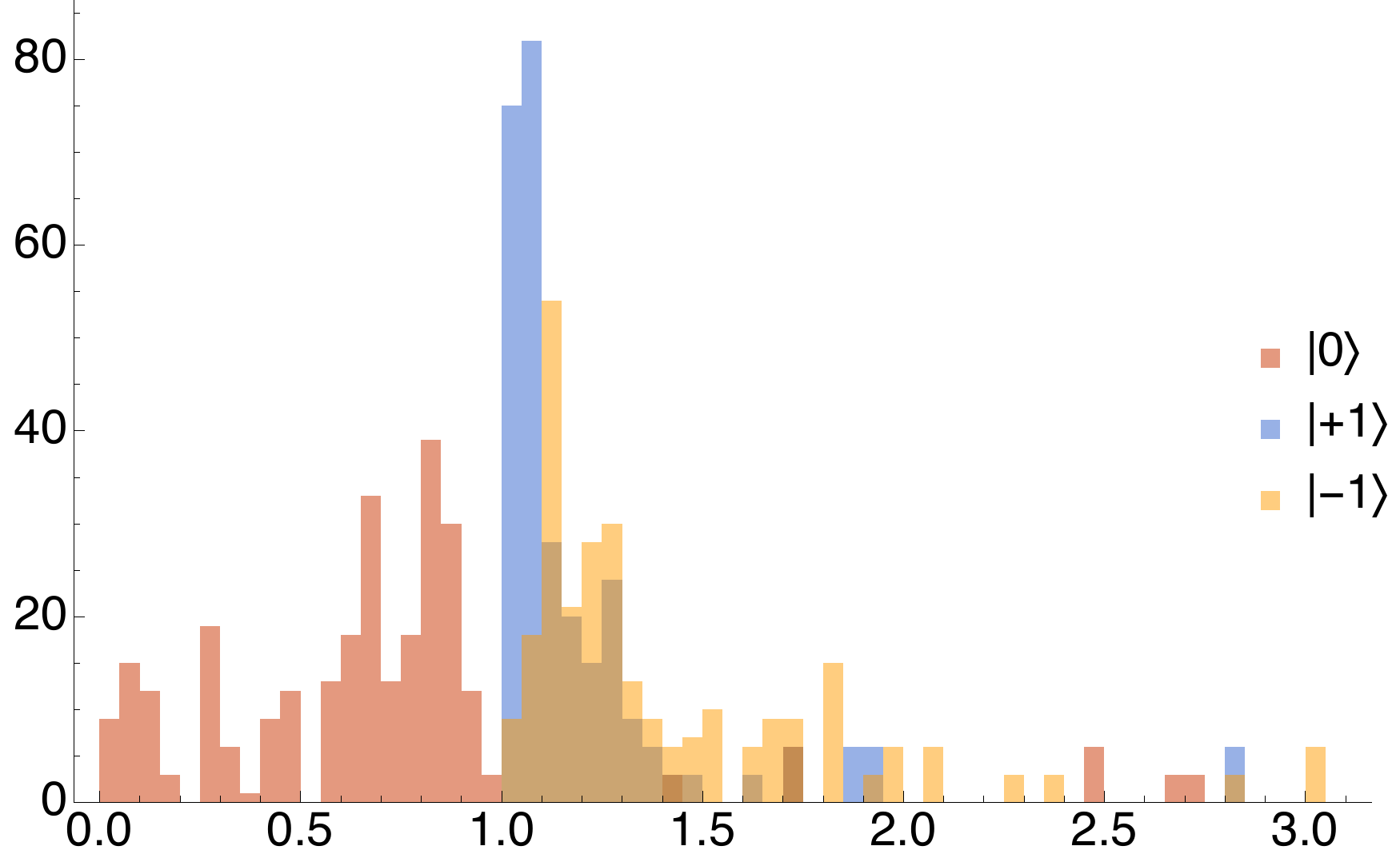} \qquad\qquad
\includegraphics[width=0.35\textwidth]{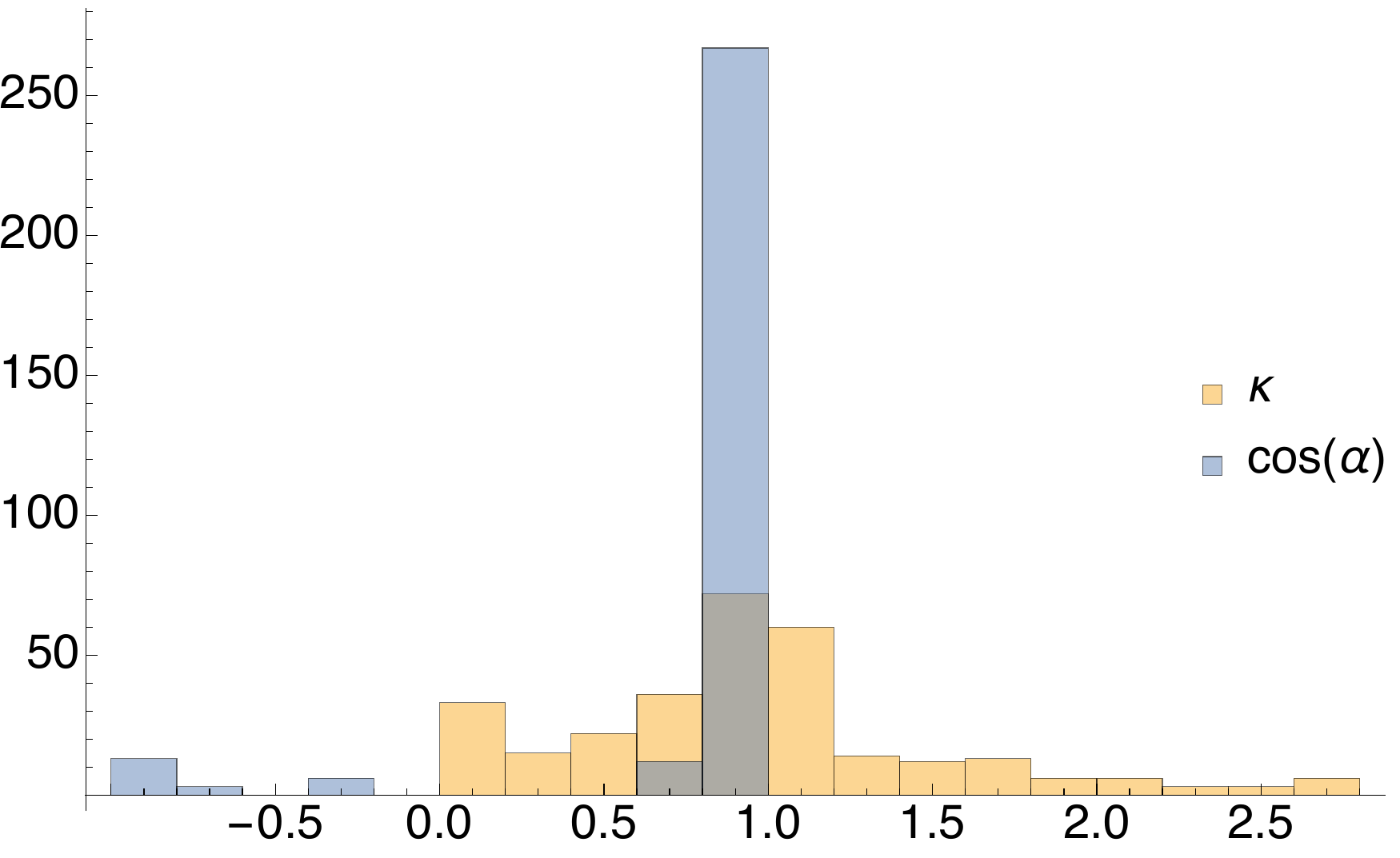}
\caption{Left: Histogram of Rabi enhancement factors (in absolute value) for the closest nuclear spins to a NV center in diamond at $B_0=500$G. While a few spins have large enhancement $>3$ (not plotted), the majority of spins have factors $1-1.5$. Right: Histogram of the relevant parameters for time-optimal control for the closest nuclear spins calculated from their coupling to a NV center in diamond at $B_0=500$G.}
\label{fig:RabiEnh}
\end{figure}

While in the main text we neglected the time required to apply $\pi$-pulses on the NV center, this time can become substantial if the number of required pulses grows. In addition, pulse errors might also accumulate and degrade the nuclear spin unitary fidelity. The actuator sequence length is thus a very important parameter, and we thus survey in Figure~\ref{fig:SeqLength} its spread over the nuclear spins of interest.
\begin{figure}[h]
\centering
\includegraphics[width=0.35\textwidth]{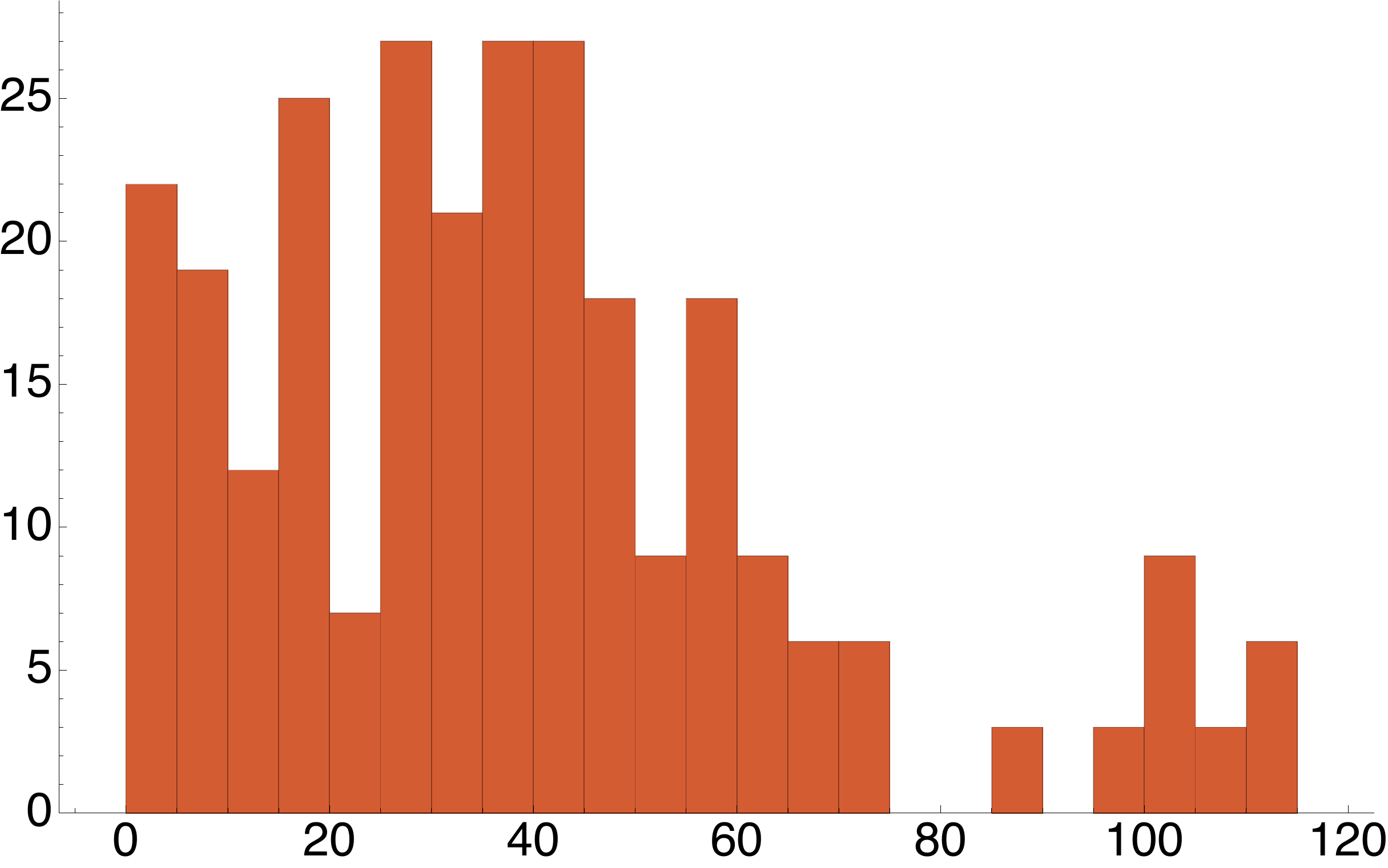} 
\caption{Number of switches required for the time optimal solution. Here we survey the closest nuclear spins to a NV center in diamond at $B_0=500$G.}
\label{fig:SeqLength}
\end{figure}

The simplest scheme to obtain rotations of the target qubit is by alternating its evolution about the two non-paralell axes for equal amounts of time. While this scheme has advantages, in particular when one also seek to preserve the coherence of the quantum actuator or when the exact rotation axes are not known with enough precision, it provides high fidelity gates only for small angles $\alpha$. In addition, the rotations are not time-optimal. In figure~\ref{fig:EqualTime} we compare the equal-time sequences with the time-optimal sequences. While the time-optimal construction can achieve in principle perfect fidelity (and we set the infidelity to $10^{-10}$ in the numerical searches) the equal-time decomposition does not leave enough degrees of freedom to achieve the desired gate. The fidelity is worse for large angles between the rotation axes and a large mismatch between the two rotation rates. When the equal time decomposition achieves acceptable fidelities, this is paid for by long decomposition times.
\begin{figure}[h]
\centering
\includegraphics[width=0.53\textwidth]{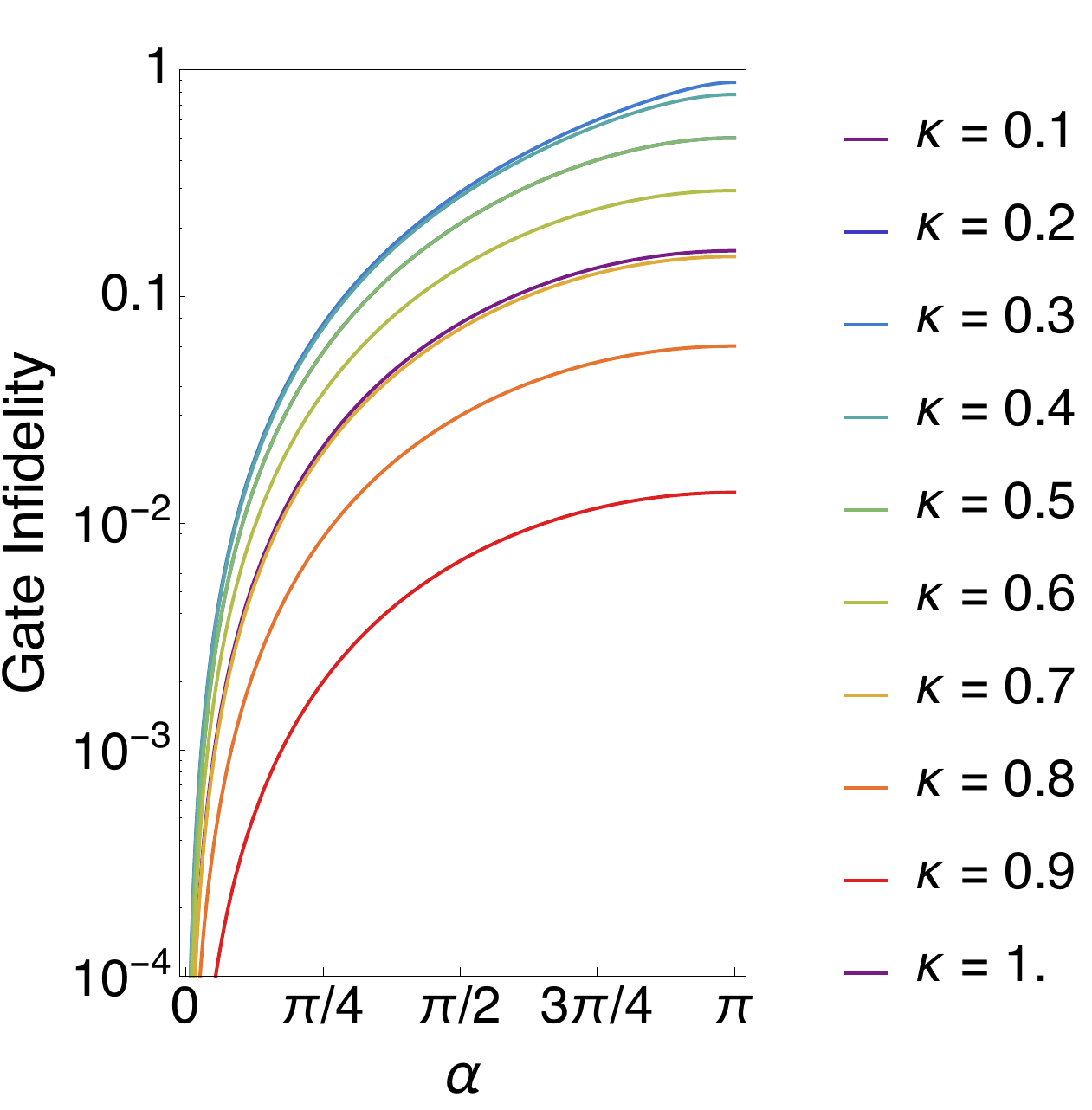} \qquad\qquad
\includegraphics[width=0.35\textwidth]{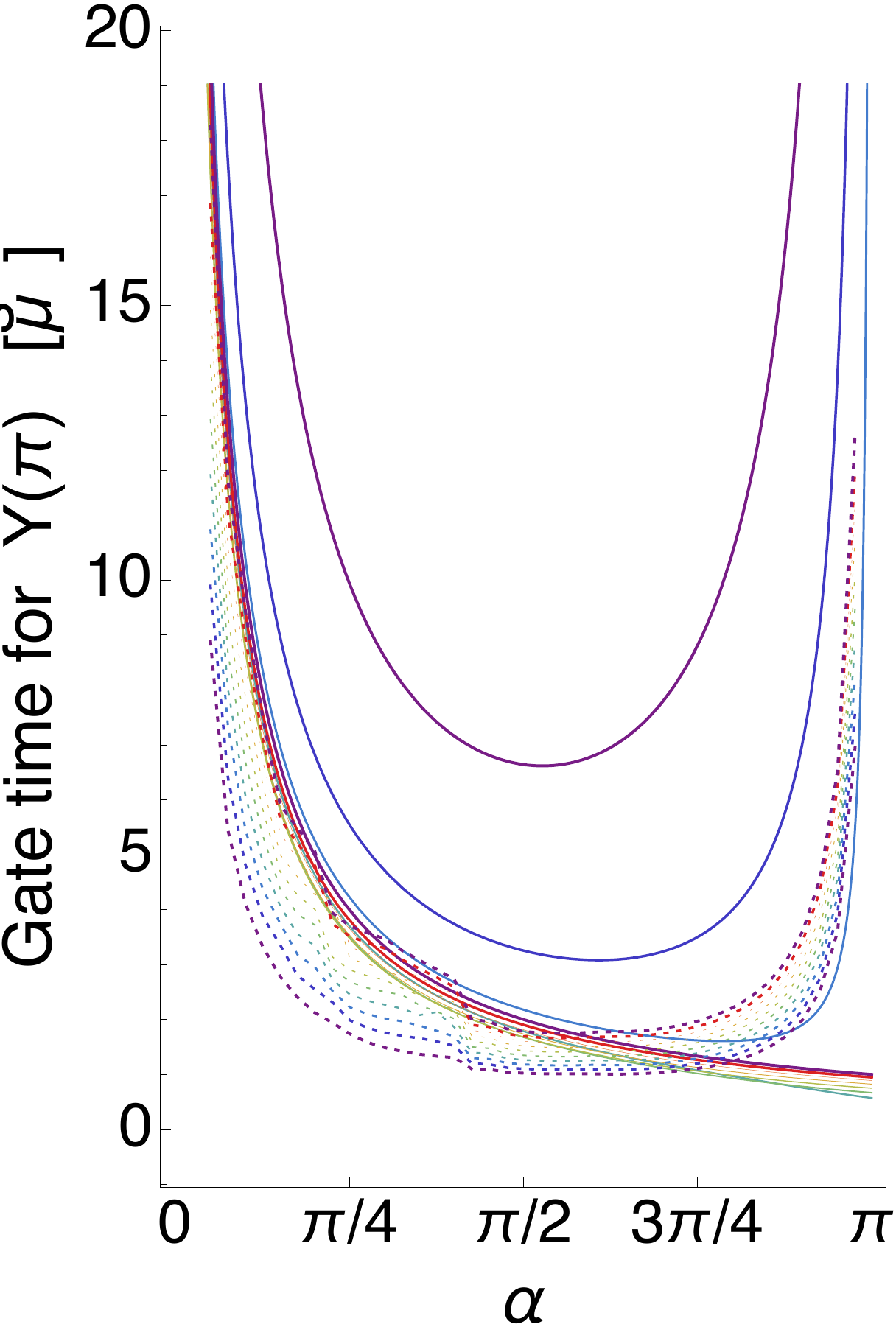}
\caption{Left: Gate Infidelity $1-|\tr{U_{eq}U_g^\dag}|$, where $U_g$ is a $\pi$ rotation about $\mathsf{Y}$. 
Here $U_{eq}$ is obtained by rotations around alternating axes (separated by an angle $\alpha$) for equal time periods. While the fidelity is good for small $\alpha$, it becomes poor at larger $\alpha$. Note that the infidelity for the time-optimal scheme is in principle 0 and was set to $<10^{-10}$ in the numerical searches. Right: Gate time for the same gate (solid lines) compared to the time-optimal solution time (dotted lines). Note that the equal-time solutions seem to be time-favorable at high $\alpha$, but then their fidelity is poor.}
\label{fig:EqualTime}
\end{figure}
\end{widetext}

\end{document}